\documentclass[twocolumn,showpacs,preprintnumbers,amsmath,amssymb,superscriptaddress,pre]{revtex4}
\usepackage{graphicx}% Include figure files
\usepackage{dcolumn}% Align table columns on decimal point
\usepackage{bm}% bold math

\usepackage{euscript,amssymb,amsmath,mathdots}

\newcommand{\be}[1]{\begin{equation}\label{#1}}
\newcommand{\ee}{\end{equation}}
\newcommand{\ba}[1]{\begin{eqnarray}\label{#1}}
\newcommand{\ea}{\end{eqnarray}}
\newcommand{\rf}[1]{(\ref{#1})}
\newcommand{\nn}{\nonumber}

\newcommand{\sign}{ \mbox{\rm sign}\,}

\def\RR{\mathbb{R}}

\def\CC{\mathbb{C}}

\def\ZZ{\mathbb{Z}}

\newcommand{\bbf}{\mathbf{f}}
\newcommand{\bg}{\mathbf{g}}

\newcommand{\cH}{\mathcal{H}}

\newcommand{\cK}{\mathcal{K}}

\newcommand{\ff}{\mathfrak{f}}

\newcommand{\fA}{\mathfrak{A}}
\newcommand{\fB}{\mathfrak{B}}

\newcommand{\fU}{\mathfrak{U}}

\begin{document}

%\preprint{arXiv}

\title{Determining role of Krein signature for three-dimensional Arnold tongues of
oscillatory dynamos}%
 %3D cones of exceptional points of the oscillatory MHD $\alpha^2$-dynamo %Force line breaks with \\

\author{Oleg N. Kirillov}

 %\altaffiliation[Also at ]{Institute of Mechanics, Moscow State Lomonosov University.}%Lines break automatically or can be forced with \\
 \email{kirillov@dyn.tu-darmstadt.de}
\affiliation{%
Technische Universit\"at Darmstadt, D-64289 Darmstadt, Germany\\
}%

\author{Uwe G\"unther}
\email{u.guenther@fzd.de}
\author{Frank Stefani}
\email{f.stefani@fzd.de}
\affiliation{%
Forschungszentrum Dresden-Rossendorf, P.O. Box 510119, D-01314 Dresden, Germany\\
                 }%

\date{\today}
%\date{April 12, 2008}% It is always \today, today,

\begin{abstract}

Using a homotopic family of boundary eigenvalue problems for the
mean-field $\alpha^2$-dynamo with helical turbulence parameter
$\alpha(r)=\alpha_0+\gamma\Delta\alpha(r)$ and homotopy parameter
$\beta \in [0,1]$, we show that the underlying network of diabolical
points for Dirichlet (idealized, $\beta=0$) boundary conditions
substantially determines the choreography of eigenvalues and thus
the character of the dynamo instability for Robin (physically
realistic, $\beta=1$) boundary conditions. In the
$(\alpha_0,\beta,\gamma)-$space the Arnold tongues of oscillatory
solutions at $\beta=1$ end up at the diabolical points for
$\beta=0$. In the vicinity of the diabolical points the space
orientation of the 3D tongues, which are cones in first-order
approximation, is determined by the Krein signature of the modes
involved in the diabolical crossings at the apexes of the cones. The
Krein space induced geometry of the resonance zones explains the
subtleties in finding $\alpha$-profiles leading to  spectral
exceptional points, which are important ingredients in recent
theories of polarity reversals of the geomagnetic field.

\end{abstract}

%may be entered using the \verb+\pacs{#1}+ command.%%


\maketitle

\section{Introduction}                                                   %!!!

Polarity reversals of the Earth's magnetic field have fascinated
geophysicists since their discovery by David and Brunhes
\cite{B1906} a century ago.
While the last reversal occurred approximately                           %!!!
780000 years ago, the mean reversal rate (averaged over the last       %!!!
few million years) is approximately 4 per Myr.                     %!!!
At least two, but very likely  three \cite{COURTILLOT}             %!!!
superchrons have been identified                                   %!!!
as ''quiet'' periods of some tens of millions of years showing    %!!!
no reversal at all.                                             %!!!

The reality of reversals is quite complex and there            %!!!
is little hope to understand all their details within a simple   %!!!
model. Recent computer simulations of the geodynamo in             %!!!
general and of reversals in particular \cite{G2002,WICHTOLSON,TAKAHASHI}  %!!!
have progressed much since the first fully coupled  3D simulations     %!!!
of a reversal by Glatzmaier and Roberts in 1995 \cite{GLRO1995}.          %!!!
Most interestingly, polarity reversals were also                    %!!!
observed in one \cite{BERHANU} of the recent liquid sodium        %!!!
dynamo experiments which have flourished during the last decade     %!!!
\cite{RMP,ZAMM}.                                                   %!!!

However, it is important to note                               %!!!
that neither in simulations
nor in experiments it is possible to accommodate all
dimensionless parameters of the geodynamo
\cite{GUBBINS}, and many of them are not even well known
\cite{ROGLA2}. In an interesting attempt to bridge the gap
of several orders of magnitude between realistic and numerically
achievable parameters, Christensen and Aubert \cite{CHRAUBERT}
were able to identify remarkable  scaling laws for some appropriate
non-dimensional numbers.                                             %!!!

The use of appropriate simplified models
\cite{MELBOURNE,HOYNG,RYAN,LUCA}
represents another attempt to understand better
the basic principle and the typical features of reversals.
Most prominent among those features  are the
distinct asymmetry (with a slow decay and a fast recovery phase)
\cite{VMG2005}, the clustering property of reversal events \cite{C2006},
and the appearance of several maxima (at multiples of ~95 kyr)
of the residence time distribution
which has been explained in terms of a stochastic resonance phenonemon
with the Milankovic cycle of the Earth's orbit excentricity
\cite{CONSOLINI,LORITO}.

One of the simplest reversal models which seems capable to explain
all those three reversal features in a consistent manner \cite{EPJB,IP}
relies basically on
the existence of an exceptional point in the spectrum of the
non-self-adjoint dynamo operator, where two real                                        %!!!
eigenvalues coalesce and continue as a complex conjugated pair of                %!!!
eigenvalues. The importance of the specific
interplay between oscillatory and non-oscillatory modes
for the reversal mechanism had been early expressed by Yoshimura
\cite{YOSHI}, Sarson and Jones \cite{SAJO}, and Gubbins and Gibbons
\cite{GUGI}.
In the framework of a simple mean-field $\alpha^2$-dynamo with
a spherically symmetric helical turbulence parameter $\alpha$
it was possible to identify reversals
as noise-triggered relaxation                                                   %!!!
oscillations in the vicinity of an exceptional point
\cite{SG2005,EPSL,S2007}.                                                       %!!!
The key point is that the exceptional point is associated with a nearby
local maximum of the                                                           %!!!
growth rate situated at a slightly lower magnetic Reynolds number.               %!!!
It is the negative slope of the growth rate curve between this local           %!!!
maximum and the exceptional point that makes stationary                       %!!!
dynamos vulnerable to  noise. Then, the instantaneous                          %!!!
eigenvalue is driven towards the exceptional point and beyond into the         %!!!
oscillatory branch where the sign change of the dipole polarity happens.       %!!!
Therefore, the existence of an exceptional point
is an essential ingredient for reversals, although non-linear
dynamics and the influence of noise must be invoked for a more
detailed understanding of those events.

From the spectral point of view, the reversal phenomenon of the
geomagnetic field is strongly linked to other fields of physics,
like van-der-Pol like oscillators \cite{vdP1926}, geometric phases
\cite{MKS2005}, $\mathcal{PT}$-symmetric quantum mechanics
\cite{Ben2007,Grae2008}, $\mathcal{PT}$-symmetric optical waveguides
\cite{PT-opt2008}, microwave resonators \cite{D2003}, and
dissipation-induced instabilities \cite{B56,M2005,K2008}.

A particular problem of all those systems in which exceptional
points are involved is a strong sensitivity of the eigenvalues on
boundary conditions (BCs).
As for the geodynamo, the periodic
occurrence of so-called superchrons is usually  attributed to the
changing thermal BCs at the core-mantle boundary \cite{COURTILLOT}, but
the growth of the inner core may also play a role \cite{S2007} by
virtue of a spectral resonance phenomenon \cite{GK2006}.

In this context it is worthwhile to note that                          %!!!
important features of dynamos are most easily understandable when     %!!!
treated with idealized (i.e. non-physical) boundary conditions.       %!!!
This was the case for explaining the famous eigenvalue symmetry        %!!!
between dipole and quadrupole modes as it was done by Proctor in 1977   %!!!
\cite{PROCTOR1,PROCTOR2}.                                               %!!!

Standing in this tradition, the present paper is devoted to a          %!!!
better understanding of the interplay of BCs, the spectral             %!!!
resonance phenomenon and oscillatory regimes in  dynamos.             %!!!

\section{Mathematical setting}                                        %!!!

The mean field MHD $\alpha^2-$dynamo \cite{krause-1} in its
kinematic regime is described by a {\em linear} induction equation
for the magnetic field. For spherically symmetric $\alpha-$profiles
$\alpha(r)$ the vector of the magnetic field is decomposed into
poloidal and toroidal components and expanded in spherical harmonics
with degree $l$ and order $m$. After additional time separation, the
induction equation reduces to a set of $l-$decoupled boundary
eigenvalue problems \cite{GK2006}, which we write in a matrix form,
convenient for the implementation of the perturbation theory
\cite{MBO97,KS04}
\be{p1}
{\bf L}\bbf:={\bf l}_0\partial_r^2 \bbf + {\bf l}_1\partial_r\bbf +
{\bf l}_2\bbf=0,\quad \fU\ff=0.
\ee
The matrices in the differential expression $\bf L$ are
\ba{p2}
{\bf l}_0&=&\left(%
\begin{array}{cc}
  1 & 0 \\
  -\alpha(r) & 1 \\
\end{array}%
\right),\quad
{\bf l}_1=\partial_r {\bf l}_0,\nn \\
{\bf l}_2&=&
\left(%
\begin{array}{cc}
  -\frac{l(l+1)}{r^2}-\lambda & \alpha(r) \\
  \alpha(r)\frac{l(l+1)}{r^2} & -\frac{l(l+1)}{r^2} -\lambda\\
\end{array}%
\right)
\ea
and $\fU:=[\fA,\fB]\in \CC^{4\times 8}$ in the BCs consists of the
blocks
\be{p3}
\fA=\left(%
\begin{array}{cccc}
  1 & 0 & 0 & 0 \\
  0 & 1 & 0 & 0 \\
  0 & 0 & 0 & 0 \\
  0 & 0 & 0 & 0 \\
\end{array}%
\right),~~
\fB=\left(%
\begin{array}{cccc}
  0 & 0 & 0 & 0 \\
  0 & 0 & 0 & 0 \\
  \beta l + 1 - \beta & 0 & \beta & 0 \\
  0 & 1 & 0 & 0 \\
\end{array}%
\right).
\ee
The vector-function $\bbf\in \tilde{\cH}=L_2(0,1)\oplus L_2(0,1)$
lives in the Hilbert space $(\tilde{\cH},(.,.))$  with inner product
$(\bbf,\bg)=\int_0^1 \overline {\bg}^T {\bbf}\, dr$, where the
overbar denotes complex conjugation, and the boundary vector $\ff$
is given as
\be{p4}
\ff^T:=\left(\bbf^T(0),\partial_r\bbf^{T}(0),\bbf^T(1),\partial_r\bbf^{T}(1)\right)\in
\CC^{8}.
\ee

We assume that $\alpha(r):=\alpha_0+\gamma\Delta\alpha(r)$, where
$\Delta\alpha(r)$ is a smooth real function
%$C^2(0,1)\ni\Delta\alpha(r):(0,1) \to \ \RR $%
with $\int_0^1 \Delta\alpha(r)
dr =0$. For a fixed $\Delta\alpha(r)$ the differential expression
$\bf L$ depends on the parameters $\alpha_0$ and $\gamma$, while
$\beta$ interpolates between idealized ($\beta=0$) BCs,
corresponding to an infinitely conducting exterior, and
physically realistic ones ($\beta=1$) corresponding to a
non-conducting exterior of the dynamo region  \cite{krause-1}.

    \begin{figure}
    \begin{center}
    \includegraphics[angle=0, width=0.48\textwidth]{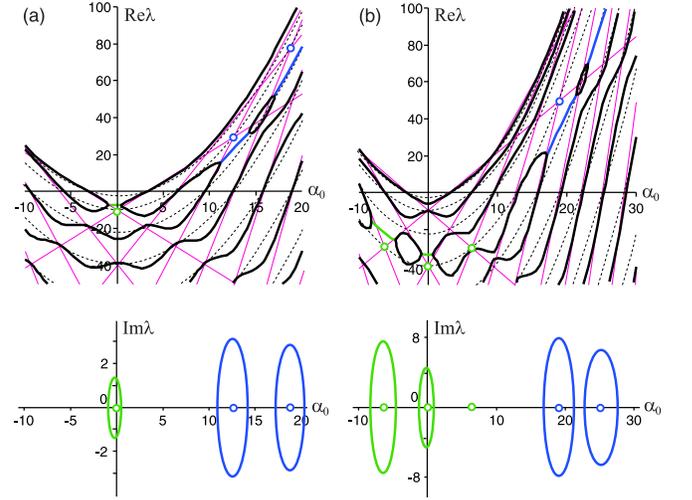}
    \end{center}
\caption{$l=0$:  (pink) spectral mesh \rf{p11} for $\gamma=0$,
$\beta=0$;\\ (dashed) eigenvalue parabolas \rf{p12} for $\gamma=0$,
$\beta=1$;\\ (black) eigenvalue branches for $\beta=0.3$,
$\Delta\alpha(r)=\cos(2\pi kr)$, and (a) $k=1$, $\gamma=2.5$, (b)
$k=2$, $\gamma=3$, with resonant overlaps near the locations of the
diabolically crossed modes having (blue) the same and (green)
different Krein signature.}
    \label{fig1}
    \end{figure}

The spectral problem \rf{p1} is not selfadjoint in a Hilbert
space, but in case of idealized BCs $(\beta=0)$ the fundamental
symmetry of the differential expression \cite{GS03,GK2006}
\be{p5}
{\bf L}_0:={\bf L}(\lambda=0)={\bf J}{\bf L}_0^\dagger {\bf J},\qquad
{\bf J}=\left(\begin{array}{cc}0 & 1\\ 1 & 0
 \end{array}\right),
\ee
makes ${\bf L}_0$ selfadjoint in a Krein space $(\cK,[.,.])$
\cite{LT} with indefinite inner product $[.,.]=({\bf J}.,.)$:
\be{p6}
[{\bf L}_0 {\bbf},{\bg}]=[{\bbf},{\bf L}_0 {\bg}], \qquad {\bbf
},{\bg} \in \cK\, .
\ee
For $\beta\neq 0$ the operator ${\bf L}_0$ is not self\-ad\-joint
even in a Krein space.

\section{From diabolic to exceptional points}

In case of constant $\alpha-$profiles $\alpha(r)\equiv
\alpha_0=$const and $\beta=0$ the spectrum and the eigenvectors of
the operator matrix ${\bf L}_0$  are \cite{GK2006}
\ba{p7}
\lambda_n^{\varepsilon}&=&\lambda_n^{\varepsilon}(\alpha_0)=-\rho_n+\varepsilon \alpha_0
\sqrt{\rho_n}\in \RR\, ,\quad \varepsilon=\pm,\nn\\
{\bbf}_n^\varepsilon&=&\left(
\begin{array}{c}
  1 \\
  \varepsilon \sqrt{\rho_n} \\
\end{array}
\right)f_n\in \RR^2\otimes L_2(0,1)\,, \quad n\in \ZZ^+
\ea
with $f_n(r)$ being normalized Riccati-Bessel functions
\be{p8}
f_n(r)=\frac{(2r)^{1/2}J_{l+\frac 12}(\sqrt{\rho_n}r)}{ |J_{l+\frac
32}(\sqrt{\rho_n})|},\quad (f_{n'},f_n)=\delta_{{n'}n}
\ee
and $\rho_n>0$  the squares of Bessel function zeros
\be{p9}
J_{l+\frac{1}{2}}(\sqrt{\rho_n})=0, \qquad
0<\sqrt{\rho_1}<\sqrt{\rho_2}<\cdots \, .
\ee
The eigenvectors $\bbf_n^+,\bbf_n^-\in \cK_\pm\subset \cK\,$
correspond to Krein space states of positive and negative signature
$\varepsilon=\pm$
\be{p10}
[\bbf_{n'}^\pm,\bbf_n^\pm ]=\pm 2 \sqrt{\rho_n}\delta_{{n'}n},~~
[\bbf_{n'}^\pm,\bbf_n^\mp]=0\, .
\ee
The spectral branches $\lambda_n^\pm$ are real-valued linear
functions of the parameter  $\alpha_0$ with signature-defined slopes
$\pm \sqrt{\rho_n}$ and form for all $l=0,1,2,\cdots$ a mesh-like
structure in the $(\alpha_0,\Re \lambda)-$plane. Spectral meshes for
neighboring mode numbers $l$ and $l+1$ have only slightly different
slopes of their branches and behave qualitatively similar under
perturbations \cite{GK2006}. Therefore basic spectral structures for
$l=1,2$ dipole and quadrupole modes can be illustrated by the
simpler but unphysical $l=0$ monopole modes which are given in terms
of trigonometric functions. The $(l=0)-$mesh built from
$\rho_n=\pi^2 n^2$ is depicted as pink lines in Fig. \ref{fig1}.

The intersection of two branches
$\lambda_{n'}^\delta,\lambda_n^\varepsilon$ with $n\neq {n'}$ occurs
at points  $(\alpha_0^{(\nu)},\lambda^{(\nu)})$ with
\be{p11}
\alpha_0^{(\nu)} :=\varepsilon \sqrt{\rho_n}+\delta
\sqrt{\rho_{n'}},\quad
\lambda^{(\nu)}:=\sigma^{(\nu)}\sqrt{\rho_n\rho_{n'}},\quad
\sigma^{(\nu)}:=\varepsilon \delta
\ee
and corresponds to double eigenvalues
$\lambda^{(\nu)}=\lambda_n^{\varepsilon}=\lambda_{n'}^{\delta}$ with
two linearly independent eigenvectors $\bbf_n^{\varepsilon}$ and
$\bbf_{n'}^{\delta}$, i.e. to so-called semi-simple eigenvalues
(diabolical points, DPs) of algebraic and geometric multiplicity two
\cite{KS04,GK2006}.

For the  $(l=0)-$mesh the diabolical crossings of the $(\varepsilon
n)$th and $(\varepsilon n + j)$th modes with the same fixed $|j|\in
\mathbb{Z^+}$ are located on a parabolic curve \cite{GK2006}
\be{p12}
\lambda(\alpha_0)=\frac{1}{4}\left(\alpha_0^2-\pi^2j^2 \right)
\ee
where $\alpha_0=\alpha_0^{(\nu)}=\pi(2n\varepsilon+j)$ and
$\lambda^{(\nu)}=\lambda(\alpha_0^{(\nu)})=\pi^2 n(n+\varepsilon
j)$. Open circles in Fig.~\ref{fig1}(a,b) indicate DPs on the
parabolas $|j|=2$  and $|j|=4$. The Krein signatures \cite{K50} of the
intersecting branches define the intersection index
$\sigma^{(\nu)}=\varepsilon\delta=\sign(\lambda^{(\nu)})$ in eqs.
\rf{p11}. Branches of different signature $\delta\neq \varepsilon$
intersect for both signs of $\alpha_0$ at $\lambda^{(\nu)} <0$
(green circles in Fig. \ref{fig1}), whereas intersections at
$\lambda^{(\nu)}
>0$  are induced by spectral branches of coinciding signatures:
for $\varepsilon=\delta=+$ at $\alpha_0>0$,
and for $\varepsilon=\delta=-$ at $\alpha_0<0$ (blue circles in
Fig.~\ref{fig1}).

For $\gamma=0$ and constant $\alpha(r)=\alpha_0$, the spectrum
remains purely real on the full homotopic family $\beta\in [0,1]$
and passes smoothly in the $(\alpha_0, \Re\lambda)-$plane from the
spectral mesh at $\beta=0$ to non-intersecting branches of simple
real eigenvalues for models with physically realistic BCs at
$\beta=1$. For the monopole model $l=0$ the full spectral homotopy
is described by the characteristic equation $
(1-\beta)\eta\left[\cos\left(\eta\right)-\cos\left(\alpha_0\right)\right]
+ 2\beta\lambda\sin\left(\eta\right)=0, $ where
$\eta(\alpha_0,\lambda)=\sqrt{\alpha_0^2-4\lambda}$, which for
physically realistic BCs $(\beta=1)$ leads to a spectrum consisting
of the countably infinite set of parabolas \rf{p12} labelled by the
index $j\in\ZZ^+$ and depicted in Fig. \ref{fig1} as dashed lines.
The reason for the $(\beta=0)-$DPs \rf{p11} to be located on the
$(\beta=1)-$parabolas \rf{p12} is that the loci of the DPs are fixed
points of the homotopy  $\forall
\beta\in[0,1]$ --- a phenomenon which indicates on their `deep imprint' in the
boundary eigenvalue problem \rf{p1}.

The eigenvalue branches with $\Re\lambda>0,\Im \lambda\neq0$
(important for the reversal mechanism \cite{SG2005,S2007}) can be
induced by deforming the constant $\alpha-$profile into an
inhomogeneous one, $\alpha(r)=\alpha_0+\gamma\Delta\alpha(r)$, with
simultaneous variation of the BCs. This process is governed by a
strong resonant correlation between the Fourier mode number of the
inhomogeneous $\Delta\alpha(r)$ and the parabola index $|j|$. This
is numerically demonstrated in Fig. \ref{fig1} (black branches) for
$\Delta\alpha(r)=\cos(2\pi k r)$ which highly selectively induces
complex eigenvalue segments  in the vicinity of DPs located on the
parabola \rf{p12} with index $j=2k$.

    \begin{figure}
    \begin{center}
    \includegraphics[angle=0, width=0.49\textwidth]{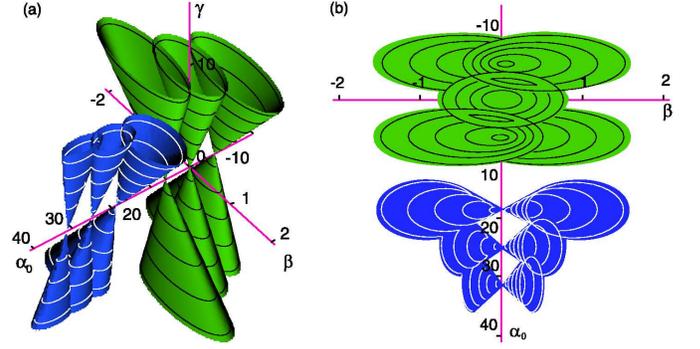}
    \end{center}
\caption{$l=0$, $\Delta\alpha(r)=\cos(4\pi r)$: (a) linear
approximation of the 3D Arnold tongues and (b) their projection
onto the $(\alpha_0,\beta)-$plane  indicating the influence of the
intersection index $\sigma^{(\nu)}$ on the inclination of the
cones.}
    \label{fig2}
    \end{figure}

The underlying influence of 'hidden' DPs on real-to-complex
transitions of the spectral branches can be made transparent by
analyzing the perturbative unfolding of the DPs \cite{KS04} at the
mesh-nodes $(\alpha_0^{(\nu)},\lambda^{(\nu)})$ under variation of
the parameters $\alpha_0$, $\beta$, and $\gamma$. In first-order
approximation this gives for the  $(l=0)-$model
\be{p15}
\lambda=\lambda^{(\nu)}-\lambda^{(\nu)}
\beta+\frac{\alpha_0^{(\nu)}}{2}(\alpha_0-\alpha_0^{(\nu)}) \pm
\frac{\pi}{2} \sqrt{D},
\ee
where $\alpha_0^{(\nu)}=\pi(\varepsilon n +\delta {n'})$,
$\lambda^{(\nu)}=\varepsilon\delta\pi^2{n'}n$, and
\ba{p16}
D&:=&\left[(\varepsilon n-\delta
{n'})\left(\alpha_0-\alpha_0^{(\nu)}\right) \right]^2
\nn
\\ &+&
{n'}n\left[(\varepsilon1+\delta1)\gamma A -(-1)^{n+{n'}}(n+{n'})\beta\pi\right]^2\nn \\
&-& {n'}n\left[(\varepsilon1-\delta1)\gamma A
-(-1)^{n-{n'}}(n-{n'})\beta\pi\right]^2
\ea
with $ A:=\int_0^1 \Delta \alpha (r) \cos[(\varepsilon n - \delta
{n'})\pi r]dr. $

For $\gamma=0$ it holds $D\ge0$, confirming that the eigenvalues
remain real under variation of the parameters $\alpha_0$ and $\beta$
only. If, additionally, $\alpha_0=\alpha_0^{\nu}$, then one of the
two simple eigenvalues \rf{p15} remains fixed under first-order
perturbations with respect to  $\beta$: $\lambda=\lambda^{(\nu)}$ in
full accordance with the fixed point nature of the DP loci under the
$\beta-$homotopy. The sign of the first-order increment of the other
eigenvalue $\lambda=\lambda^{(\nu)}-2\lambda^{(\nu)}\beta$ depends
on the sign of $\lambda^{(\nu)}$ and, therefore, via \rf{p11}
directly on the Krein signature of the modes involved in the
crossing $(\alpha_0^{(\nu)},\lambda^{(\nu)})$.

In general, there exist parameter combinations yielding $D<0$ and
thus creating complex eigenvalues. Eq. \rf{p16} implies that in
first-order approximation the domain of oscillatory solutions with
${\Re}\lambda\ne0$ and ${\Im}\lambda\ne0$ in the $(\alpha_0,
\beta,\gamma)$-space is bounded by the
conical surfaces $D=0$ with  apexes at the DPs
$(\alpha_0^{(\nu)},0,0)$, as shown in Fig. \ref{fig2}.
Such domains, especially in case of $r-$periodic $\alpha-$profiles,
are in fact Arnold tongues corresponding to zones of parametric
resonance \cite{K50} in Mathieu-type equations whose analysis in \cite{A83}  was motivated  just by Zeldovich's studies on MHD dynamos.

At the boundary $D=0$ the eigenvalues are twofold degenerate and
non-derogatory, that is they have Jordan chains consisting of an
eigenvector and an associated vector. Thus, DPs in the $(\alpha_0,
\beta,\gamma)$-space unfold into 3D conical surfaces
consisting of  exceptional points (EPs).

The conical zones develop according to resonance selection rules
similar to those discovered in \cite{GK2006} for the case $\beta=0$.
For example, with $\Delta \alpha(r)=\cos(2\pi k r)$,
$k\in\mathbb{Z}$, the constant $A$ in \rf{p16} yields
\be{p18}
A= \left \{\begin{array}{r}
1/2,\quad 2k=\varepsilon n- \delta {n'} \\
0,\quad 2k\ne\varepsilon n- \delta {n'}
\end{array}
 \right.
\ee
so that in  first-order approximation only DPs located on the
$(j=2|k|)-$parabola \rf{p12} show a DP-EP unfolding (in accordance
with numerical results in Fig. \ref{fig1}). The cone apexes
correspond to $2|k|-1$ DPs with negative intersection index \rf{p11}
$\sigma^{(\nu)}=-$ and countably infinite DPs with
$\sigma^{(\nu)}=+$. The two groups are shown in Fig. \ref{fig2} in
green and blue, respectively.

The real parts of the perturbed eigenvalues are given by $\Re
\lambda=\lambda^{(\nu)}(1-
\beta)+\alpha_0^{(\nu)}\left(\alpha_0-\alpha_0^{(\nu)}\right)/2$
and for fixed $\alpha_0$ and increasing $\beta$ they are shifted
(for both groups) away from the original DP positions toward the
$(\Re\lambda=0)-$axis (cf. the numerical results in Fig. \ref{fig1})
--- an effect which is similar to the
self-tuning mechanism of field-reversals uncovered in \cite{SG2005}.

Apart from this similarity, the eigenvalues of the two cone-groups
show significant differences.  The $2|k|-1$ cones of the first
group have non-trivial intersection with the plane $\beta=0$. In
this $(\alpha_0,\gamma)-$plane the zones of decaying oscillatory
modes are $\gamma \rightleftharpoons-\gamma $ symmetric and
defined by the inequality
\be{p19} \left(\alpha_0 \pm 2\pi(n-|k|) \right)^2<
\frac{\gamma^2}{4}\left[ 1- \left(\frac{n-|k|}{|k|}
\right)^2\right],
\ee
where $n=1, 2,\ldots,|k|$. For $k=2$ there are  three primary Arnold
tongues: $4\alpha_0^2<\gamma^2$ and $16\left(\alpha_0 \pm 2\pi
\right)^2<3\gamma^2$.

    \begin{figure}
    \begin{center}
    \includegraphics[angle=0, width=0.49\textwidth]{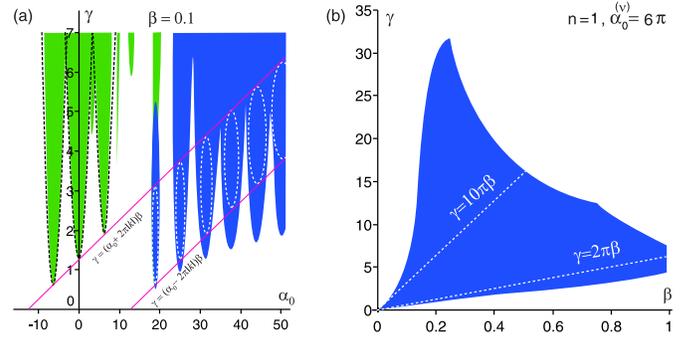}
    \end{center}
\caption{$l=0$: numerically calculated Arnold tongues for
$\Delta\alpha(r)=\cos(2\pi k r)$, $k=2$, and $\lambda^{(\nu)}<0$
(green) or $\lambda^{(\nu)}>0$ (blue) and their approximations
(dashed lines) (a) in the $(\alpha_0,\gamma)-$plane  and (b) in the
$(\beta,\gamma)-$plane.}
    \label{fig3}
    \end{figure}

The cones of the second group meet the plane $\beta=0$ only at the
apexes, having their skirts located in the sectors
$[\beta>0,\gamma\,\sign(\alpha_0)>0]$ and
$[\beta<0,\gamma\,\sign(\alpha_0)<0]$ (cf. Fig. \ref{fig2}(b)).
Therefore, in models with idealized BCs $(\beta=0)$ complex
eigenvalues occur only in zones \rf{p19} in the
$(\alpha_0,\gamma)$-plane.

The different oscillatory behavior induced by the two cone groups
has its origin in the different Krein-signature defined inclination
of the $(D<0)-$cones with respect to the $(\beta=0)-$plane.

Passing from $\beta=0$ to a parallel $(\beta\neq0)-$plane, the
$(D<0)-$tongues \rf{p19}, corresponding to $\lambda^{(\nu)}<0$
deform into cross-sections bounded by hyperbolic curves (black
dashed lines in Fig.~\ref{fig3}(a))
\ba{p20}
-4k^2(\alpha_0\pm2\pi(n{-}|k|))^2 &+& n(2|k|{-}n)(\gamma\pm 2\pi(n{-}|k|)\beta)^2\nn\\&>&n(2|k|{-}n)4\pi^2\beta^2k^2,
\ea
with $n=1, 2,\ldots,|k|$. Since $n\le|k|$, the lines $\gamma=\pm2\pi
n\beta$ and $\gamma=\pm2\pi (n-2|k|)\beta$, bounding the
cross-sections of the 3D cones by the plane
$\alpha_0^{(\nu)}=\pm2\pi(n-|k|)$, always have the slopes of
different sign. This allows decaying oscillatory modes for $\beta=0$
due to variation of $\gamma$ only.

The $(\beta \neq 0)-$cross-sections of the cones with
$(\lambda^{(\nu)}>0)-$apexes have the form of ellipses (white dashed
lines in Fig.~\ref{fig3}(a))
\ba{p21}
~~4k^2\left(\alpha_0 \pm 2\pi(n{+}|k|) \right)^2&{+}&n(2|k|{+}n)
\left(\gamma \pm 2\pi(n{+}|k|)\beta \right)^2\nn \\ &<&n(2|k|{+}n)4\pi^2\beta^2k^2,
\ea
where $n=1,2,\ldots$.  In the $(\beta\neq 0)-$plane the ellipses are
located inside the stripe with  boundaries
$\gamma=(\alpha_0\pm2\pi|k|)\beta$ (pink lines in
Fig.~\ref{fig3}(a)), while the hyperbolas lie outside this stripe.
Moreover, since in the plane $\alpha_0^{(\nu)}=\pm2\pi(n+|k|)$ the
boundary lines $\gamma=\pm2\pi n\beta$ and $\gamma=\pm2\pi
(n+2|k|)\beta$ have slopes of the same sign, the $\gamma$-axis does
not belong to the instability domains, showing that for growing
oscillatory modes the parameters $\beta$ and $\gamma$ have to be
taken in a prescribed proportion, see Fig.~\ref{fig3}(b).

The amplitude $\gamma$ of the inhomogeneous perturbation of the
$\alpha$-profile $\gamma\Delta\alpha(r)$ is limited both from below
and from above in the vicinity of the DPs with $\sigma^{(\nu)}>0$.
However, numerical calculations indicate that this property can
persist on the whole interval $\beta\in[0,1]$,  see
Fig.~\ref{fig3}(b), in agreement with the earlier findings of
\cite{SG2003}.

\section{Conclusions}                                                  %!!!

In summary, we have found that the underlying network of DPs and
their intersection indices for
$\beta=0$ substantially determine the choreography of eigenvalues
for $\beta=1$ and, in particular, the loci of EPs which are
important to explain the reversals of the geomagnetic field.
Although this has been exemplified for the unphysical monopole ($l=0$)   %!!!
mode of a simplified spherically symmetric $\alpha^2$ dynamo model,      %!!!
the general idea is well generalizable to physical modes and to more      %!!!
realistic dynamo models. Work in this direction is in progress.         %!!!
                      %!!!

\begin{acknowledgments}
The research of O.N.K. was supported by DFG  grant  HA 1060/43-1
as well as by the Saxon Ministry of Science grant
4-7531.50-04-844-07/8 and that of U.G. and F.S. by DFG
Sonderforschungs\-be\-reich 609.
\end{acknowledgments}
%and which allows for the conclusion that models with idealized BCs
%can provide further insight even into the realistic polarity
%reversal regimes of \cite{SG2005}.

%\bibliography{kgs}

\end{document}